\documentclass[aps,prd,preprint,groupedaddress,showpacs]{revtex4}
\usepackage{epsf,epsfig,graphicx,color,hyperref}
\newcommand{\gapr}{\gamma^\prime}
\newcommand{\geff}{g_{\rm eff}}
\newcommand{\geffsr}{g^{1/2}_{\rm eff}}
\newcommand{\heff}{h_{\rm eff}}
\newcommand{\gastsr}{g^{1/2}_{\ast}}
\newcommand{\tdpdec}{T_{\gapr {\rm dec}}}
\newcommand{\dpdec}{\gapr {\rm  dec}}

\begin{document}

\title{Dark Photon as Fractional Cosmic Neutrino Masquerader}

\author{Kin-Wang Ng}
\email{nkw@phys.sinica.edu.tw}
\affiliation{Institute of Physics, Academia Sinica, Taipei, Taiwan 11529, R.O.C.}
\author{Huitzu Tu}
\email{huitzu@phys.sinica.edu.tw}
\affiliation{Institute of Physics, Academia Sinica, Taipei, Taiwan 11529, R.O.C.}
\author{Tzu-Chiang Yuan}
\email{tcyuan@phys.sinica.edu.tw}
\affiliation{Institute of Physics, Academia Sinica, Taipei, Taiwan 11529, R.O.C.}

\begin{abstract}

Recently, Weinberg proposed a Higgs portal model with a spontaneously broken global $U(1)$ symmetry 
in which Goldstone bosons may be masquerading as fractional cosmic neutrinos. 
We extend the model by gauging the $U(1)$ symmetry. 
This gives rise to the so-called dark photon and dark Higgs.
The dark photons can constitute about 0.912 (0.167) to the effective 
number of light neutrino species if they decouple from the thermal bath 
before the pions become non-relativistic and
after (before) the QCD  transition.
Restriction on the parameter space of the portal coupling and the dark Higgs mass is obtained 
from the freeze-out condition of the dark photons. 
Combining with the collider data constraints on the invisible width of the standard model Higgs
requires the dark Higgs mass to be less than a few GeV.

\end{abstract}

\pacs{98.80.Cq, 11.15.Ex, 98.70.Vc}

\date{\today}

\maketitle

\section{Introduction}

The cosmic microwave background (CMB) radiation, if combined with other observational data, 
can be used to constrain the effective number of light neutrino species $N_{\rm eff}$.
The WMAP9 data combined with eCMB, BAO, and $H_0$ measurements has inferred 
$N_{\rm eff} = 3.55^{+0.49}_{-0.48}$ at 68$\%$ CL~\cite{Hinshaw:2012aka}.
Latest Planck data combined with WP, highL, BAO, and $H_0$ measurements gives 
$N_{\rm eff} = 3.52^{+0.48}_{-0.45}$ at 95$\%$ CL~\cite{Ade:2013zuv}.
Most recently, with the inclusion of the B-mode polarization data by the 
BICEP2 experiment~\cite{Ade:2014xna}, evidence for an extra weakly-interacting light species 
becomes more favorable, with $N_{\rm eff} \simeq 4$ (see e.g. 
Refs.~\cite{Giusarma:2014zza,Dvorkin:2014lea,Archidiacono:2014apa,Zhang:2014nta}). 
However one must be cautious about the primordial gravitational waves interpretation of the BICEP2 data since new analysis \cite{Liu:2014mpa,Mortonson:2014bja,Flauger:2014qra} has pointed out that light scattering from dust as well as synchrotron radiation produced by electrons wandering around the galactic magnetic fields within the Milky Way may also generate the B-mode in the foreground.
On the other hand, these bounds are consistent with that from the big bang nucleosynthesis (BBN) $N_{\rm eff} = 3.71^{+0.47}_{-0.45}$ (see e.g. Ref.~\cite{Steigman:2012ve}),
while the standard scenario with three active, massless neutrinos
predicts $N_{\rm eff} = 3.046$ at the CMB epoch~\cite{Mangano:2005cc}.

There has been a lot of attempts to account for a possible deviation from the theoretical prediction~\cite{Steigman:2012ve}.
Recently, Weinberg~\cite{Weinberg:2013kea} has investigated whether Goldstone bosons 
can be masquerading as fractional cosmic neutrinos. The motivation is that they would be massless or nearly massless, and
their characteristic derivative coupling would make them very weakly-interacting
at sufficiently low temperatures. The most crucial criterion is that these Goldstone bosons 
must decouple from the thermal
bath early enough so that their temperature is lower than that of the neutrinos.
To realize this idea, Weinberg has considered the simplest possible broken continuous symmetry, 
a global $U(1)$ symmetry associated with the conservation of some quantum number $W$.
A complex scalar field $\chi(x)$, which is a singlet in the Standard Model (SM) while carrying a nonvanishing value of $W$,
is introduced for breaking this symmetry spontaneously. The thermal history of the resulting Goldstone boson depends crucially on its coupling to the SM Higgs field and the mass of the radial field. It has been shown~\cite{Weinberg:2013kea} that these Goldstone bosons can contribute about $0.39$ to the effective number of light species. 
Recently, collider phenomenology of the Goldstone boson 
has been discussed~\cite{Cheung:2013oya} and  
energy loss due to Goldstone boson emission in the cooling 
of a post-collapse supernova core were examined~\cite{Keung:2013mfa}. 
Other low-energy experimental constraints on the model from dark matter experiments 
as well as $B$-meson and kaon decays into invisibles have also been 
studied in Ref.~\cite{Anchordoqui:2013bfa}. 
Implications of light sterile neutrino species are also scrutinized 
in further 
details~\cite{Giusarma:2014zza,Dvorkin:2014lea,Archidiacono:2014apa,Zhang:2014nta,Leistedt:2014sia}
after the BICEP2 experimental result was announced.

In this work we extend the Higgs portal model by gauging the $U(1)$ symmetry, thus trading the Goldstone boson with a massive gauge field, which is indeed a type of the so-called dark photon. 
If the dark photon couples to the thermal bath
as the Goldstone boson case \cite{Weinberg:2013kea} until the muon annihilation era, with its three polarization states its contribution to $N_{\rm eff}$ would be three times larger, namely $\Delta N_{\rm eff} = 3 \times 0.39 = 1.17$.
This is inconsistent with the Planck or even the combined Planck + BICEP2 data. 
The dark photon must therefore decouple at earlier era.
In this work we discuss the cosmology of the dark photon and its contribution to the effective number of light neutrino species. 
In section II, we set up the notations for the dark $U(1)$ Higgs model. 
In section III, we discuss the collider bounds on the portal coupling that connect the dark Higgs sector with the SM Higgs.
In section IV, we study the possibility of treating the light dark photons as fractional cosmic neutrinos contributing to the cosmic soup.  
Thermal production and annihilation of the dark photon are studied in section V while
the freeze-out of the dark photon near the QCD transition era is analyzed in section VI.
In section VII, the supernova bound is briefly discussed. 
We finally summarize in section VIII.

\section{The Model}

With the extra $U(1)$ gauge field added to Weinberg's Higgs portal model, the Lagrangian for the scalar fields is
\begin{equation}
  \mathcal{L}_{\rm scalar} = (D_\mu \Phi)^\dagger (D^\mu \Phi) + (D_\mu \chi)^\ast (D^\mu \chi)
  - V_{\rm scalar}\, ,
\label{Lggn}
\end{equation}
with
\begin{equation}
  V_{\rm scalar} = - \mu^2_\Phi\, \Phi^\dagger \Phi + \lambda_\Phi\, (\Phi^\dagger \Phi)^2 
  - \mu^2_\chi\, \chi^\ast \chi + \lambda_\chi\, (\chi^\ast \chi)^2 
  + \lambda_{\Phi \chi}\, (\Phi^\dagger \Phi)\, (\chi^\ast \chi)\, ,
\end{equation}
and 
\begin{equation}
   D_\mu \chi = (\partial_\mu + i g_D\, C_\mu)\, \chi\, ,
\end{equation}
where $\Phi$ is the Higgs field in the SM, $C_\mu$ is a $U(1)$ gauge field with a gauge coupling $g_D$, and $\mu$'s and $\lambda$'s are model constants. 
We will call $\lambda_{\Phi \chi}$ the portal coupling in what follows.
Note that Weinberg's Higgs portal model is given by the Lagrangian~(\ref{Lggn}) 
with $g_D=0$.

In the unitary gauge, the scalar fields are 
\begin{displaymath}
  \chi = \frac{1}{\sqrt{2}}\, \left(v_D + h_D (x) \right)\, , \hspace{0.5cm}
  \Phi = \frac{1}{\sqrt{2}}\, \left( {\begin{array}{*{20}c} 0 \\ v + h (x) \\ 
  \end{array}} \right)\, . 
\end{displaymath}
From 
\begin{equation}
   |D_\mu \chi|^2 = \frac{1}{2}\, \left[(\partial_\mu h_D)^2 + 
   g^2_D\, C^2_\mu\, (v_D + h_D (x))^2 \right]\, ,
\end{equation}
the mass of the dark photon denoted by $\gapr$ is $m_{\gapr} = g_D\, v_D$.
The two scalar fields $h (x)$ and $h_D (x)$ mix to give 
two mass eigenstates $h_1$ and $h_2$, with eigenvalues 
\begin{equation}
   m^2_{1, 2} = \frac{1}{2}\, \left[{\rm Tr} (M^2) \pm 
   \sqrt{\left( {\rm Tr} (M^2) \right)^2 - 4\, {\rm Det} (M^2)} \right]\, ,
\end{equation}
and a mixing angle 
\begin{equation}
  \sin 2 \alpha = \frac{2 \lambda_{\Phi \chi}\, v\, v_D}{m^2_1 - m^2_2}\, .
\end{equation}
Here $M^2$ is the mass-squared matrix that can be easily read off from the 
Lagrangian~(\ref{Lggn}). 
We will identify $h_1$ to be the physical SM Higgs boson $h_{\rm SM}$ with 
$m_1\simeq 125~{\rm GeV}$ and $h_2$ be the physical dark Higgs boson with 
mass $m_2$ much less than $m_1$.

In this model, the interaction of the dark photons with the SM particles arises entirely
from a mixing of the dark Higgs boson with the SM Higgs boson. 
From the interaction term $2 g^2_D\, v_D\, h_D \, C_\mu C^\mu$ arises from the covariant coupling
$\vert D_\mu \chi \vert^2$ 
and the portal mixing term $\lambda_{\Phi \chi}\, v\, v_D\, h\, h_D$, 
as well as the SM Higgs-fermion coupling $- m_f\,  h\, \bar{f} f /v$,
an effective interaction between the dark photon and any SM fermion $f$, 
\begin{equation}
   2 \lambda_{\Phi \chi}\, \frac{m_f\, m^2_{\gapr}}{m_{h_D}^2 m_h^2} \bar{f}\, f\,  
   C_\mu C^\mu\, ,
\label{ealpha}
\end{equation}
is produced. 
Here we assume $m_h \gg m_{h_D} \gg m_{f,\gamma'}$. 
The dark photon can also couple to SM gluon and photon via triangle fermion loops.
We do not consider a possible kinetic mixing between the dark photon and the SM photon
which may lead to interesting collider phenomenology such as multilepton jets as studied in
Ref.~\cite{Chang:2013lfa}.
So the model parameters are $\lambda_{\Phi \chi}$, $v_D$, $g_D$ (or $m_{\gapr} = g_D\, v_D$),
and $m_2$.
In order for the dark photons masquerading as the cosmic neutrino species and contributing 
to $N_{\rm eff}$, $m_{\gapr}$ must be in the $\lesssim$ eV range.
There is a cosmological bound to the mass of the dark photon from demanding that its 
relic density (see e.g. Ref.~\cite{Kolb:1990vq})
\begin{equation}
   \Omega_{\gapr} h^2 = 7.83 \times 10^{-2} \frac{3}{\heff (\tdpdec)}\, 
   (m /{\rm eV}) \lesssim 1\, .
\end{equation}
If the dark photon decouples from the thermal bath at 
$T_{\gapr {\rm dec}} \simeq 100~{\rm MeV}$, the bound
is $m_{\gapr} \lesssim 70~{\rm eV}$, and is even weaker for larger $T_{\gapr {\rm dec}}$.

In the unitary gauge, the Goldstone boson is completely absorbed and becomes
the longitudinal polarisation state of the massive gauge boson.
Equivalence theorem states that in the high-energy limit the Goldstone bosons will control the emission or absorption of the massive gauge bosons.
This can be seen clearly in the dark photon polarization sum 
\begin{equation}
  \sum_{\rm pol.} \epsilon^\ast_\mu (q)\, \epsilon_\nu (q) = 
  - g_{\mu \nu} + \frac{q_\mu\, q_\nu} {m^2_{\gapr}}\, ,
\end{equation}
in which the first term is the contribution from the two transverse polarization states,
while the second term that from the longitudinal polarization state.
In the scattering processes we will consider, the energies are much higher than $m_{\gapr}$,
so we expect similar results as what we obtained for the Goldstone bosons in 
Ref.~\cite{Keung:2013mfa}.

\section{Collider Bounds From Higgs Invisible Width}

The non-standard decay branching ratio of the SM Higgs boson is constrained to 
$\Gamma_{h \rightarrow {\rm inv}} < 1.2~{\rm MeV}$ (branching ratio about $22\%$)
by the results of a global fitting to the most updated data from the CMS and ATLAS 
experiments at the Large Hadron Collider (LHC), as well as those 
from the Tevatron~\cite{Cheung:2013kla,Giardino:2013bma,Ellis:2013lra,Belanger:2013xza}. 
This turns into a bound on some combination of the parameters in the gauged Weinberg's 
Higgs portal model.
In this model, the Higgs non-standard decay channels are $h \rightarrow \gapr \gapr$ and 
$h \rightarrow h_D h_D$ which can be taken as invisible modes.

The decay width for $h \rightarrow \gapr \gapr$ is 
\begin{equation}
   \Gamma_{h \rightarrow \gapr \gapr} = \frac{1}{32 \pi}\, 
   \frac{\lambda^2_{\Phi \chi} v^2}{(m^2_1 - m^2_2)^2}\, 
   \frac{\sqrt{m^2_1 - 4 m^2_{\gapr}}}{m^2_1}\, 
   \left[ m^4_1 - 4 m^2_1 m^2_{\gapr} + 12 m^4_{\gapr}\right]\, ,
\end{equation}
and that for $h \rightarrow h_D h_D$ is
\begin{equation}
   \Gamma_{h \rightarrow h_D h_D} = \frac{1}{32 \pi}\, 
   \lambda^2_{\Phi \chi} v^2\, 
   \frac{\sqrt{m^2_1 - 4 m^2_2}}{m^2_1}\, . 
\end{equation}

In the limit $m_1 \gg m_2$, $m_1 \gg m_{\gapr}$, one obtains a constraint
\begin{equation}
\label{colliderbound}
   |\lambda_{\Phi \chi}| < 0.011\, .
\end{equation}
This bound is similar to the one obtained previously for the Goldstone boson 
case \cite{Weinberg:2013kea,Cheung:2013oya} 
which should be expected by invoking the equivalence theorem.

As for future improvement, the LHC is expected to reach a sensitivity of 
$\Gamma_{h \rightarrow {\rm inv}} < 9\%$ in the year 2035.
This means that the LHC bound on the dark Higgs coupling to SM Higgs 
$|\lambda_{\Phi \chi}|$ will become $\simeq 1.56$ times stronger by then.
Furthermore, if the International Linear Collider (ILC) construction could begin 
in 2015/2016 and complete after 10 years, it may constrain the 
branching ratio of Higgs invisible decays to $< 0.4$ - $0.9\%$~\cite{Bechtle:2014ewa}
in the best scenarios.
If this can be realized, the collider bound on $|\lambda_{\Phi \chi}|$ will be 
improved by a factor of $5$ - $7$. 
Similar sensitivity of this coupling can be reached at the 240 GeV circular electron positron collider (CEPC) \cite{CEPC} Higgs factory 
proposed recently by China.

\section{Effective Number of Light Neutrino Species}

The total energy density and pressure of all particle species $j$ in kinetic equilibrium
at temperature $T_j$ can be expressed in terms of the photon temperature $T$ as
\begin{eqnarray}
   \rho &=& T^4\, \sum_{j} \left(\frac{T_j}{T} \right)^4 \frac{g_j}{2 \pi^2} 
   \int^\infty_{x_j} d u \frac{(u^2 - x^2_j )^{1/2}\, u^2}
   {e^u \pm 1}\, , \nonumber \\
   p &=& T^4\, \sum_{j} \left(\frac{T_j}{T} \right)^4 \frac{g_j}{6 \pi^2} 
   \int^\infty_{x_j} d u \frac{(u^2 - x^2_j )^{3/2}}{e^u \pm 1}\, , 
\end{eqnarray}
where $x_j \equiv m_j / T$.
In good approximation, one only need to include contributions from the
relativistic species to the energy and the entropy density
\begin{equation}
   \rho = \frac{\pi^2}{30}\, \geff (T)\, T^4\, , \hspace{1cm}   s = \frac{\rho + p}{T} = \frac{2 \pi^2}{45}\, \heff (T)\, T^3\, .
\end{equation}
The effective degrees of freedom for the energy and the entropy density are then
\begin{equation}
   \geff (T) = \sum_i C_i\, g_i\, \left( \frac{T_i}{T} \right)^4\, , 
   \hspace{1.5cm} \heff (T) = \sum_i C_i\, g_i\, \left( \frac{T_i}{T} \right)^3\, , 
\end{equation}
respectively, with $C_i = 1$ for $i=$ boson, and $\frac{7}{8}$ for $i=$ fermion.
The evolution of $\geff (T)$ and $\heff (T)$ has been calculated in 
Refs.~\cite{Olive:1980wz,Srednicki:1988ce}, and the problem of correct matching the 
degrees of freedom between the low and high temperature regions was studied in 
Ref.~\cite{Hindmarsh:2005ix}.
Conventionally, the contribution of neutrinos is parametrized by the effective number of 
light neutrino species, $N_{\rm eff}$, via the relation
\begin{equation}
   \rho = \left[1 + \frac{7}{8}\, \left(\frac{T_\nu}{T} \right)^4\, N_{\rm eff} 
   \right]\, \rho_\gamma\, .
\end{equation}
This definition can also accommodate any exotic light species $X$.
Its temperature relative to the neutrino temperature $T_\nu$ is determined by the time it decouples from the thermal bath.
Conservation of the entropy per comoving volume dictates that  
the temperature of the Universe evolves as 
$T \propto \heff^{-1/3}\, a^{-1}$ with the scale factor $a$.
When a particle species becomes non-relativistic, $\heff$ decreases. 
The entropy of this particles species is then transferred to the other relativistic particle
species remaining in the thermal bath.
On the other hand, massless particles that are decoupled from the thermal bath will not share in this entropy transfer. Its temperature simply scales as $T_i \propto a^{-1}$.
Therefore, after the $e^+ e^- \rightarrow \gamma \gamma$ annihilation, the temperature 
of the neutrinos is lower than that of the photon by $T_\nu = (4/11)^{1/3}\, T$.
Now suppose the light species $X$ decouples from the thermal bath at an 
ealier epoch than the neutrinos.
After the neutrino decoupling, its temperature relative to the neutrino temperature 
is then fixed at
\begin{equation}
   \frac{T_X}{T_\nu} = \left(\frac{\heff (T_{\nu\, {\rm dec}})}
   {\heff (T_{X\, {\rm dec}})} \right)^{1/3}\, .
\end{equation}
Here $\heff (T_{\nu\, {\rm dec}}) =10.75$ is the SM $\heff$ value at 
neutrino decoupling, and $\heff (T_{X\, {\rm dec}})$ that at the $X$ species decoupling.
The CMB data thus impose a constraint on the property of the $X$ particle,
\begin{equation}
   C_X\, g_X\, \left( \frac{T_X}{T_{\rm CMB}} \right)^4
   \leq \frac{7}{8} \cdot 2\, (N^{\rm CMB}_{\rm eff} - 3.046)\, 
    \left( \frac{4}{11} \right)^{4/3}\, ,
\end{equation}
if $m_X \lesssim T_{\rm CMB} \sim 1~{\rm eV}$, 
where $N^{\rm CMB}_{\rm eff}$ is the CMB upper bound on $N_{\rm eff}$.
The dark photon with $C_X=1$ for boson and $g_X=3$ due to the three polarization states would then contribute to $N_{\rm eff}$ with
\begin{equation}
   \Delta N^{\gapr}_{\rm eff} = \frac{4}{7}\, \cdot 3\, 
   \left(\frac{10.75}{\heff (T_{\gapr\, {\rm dec}})} \right)^{4/3}\, .
\end{equation}

\begin{center}
\begin{figure}[h!]
  \includegraphics[width=0.6\textwidth,angle=-90]{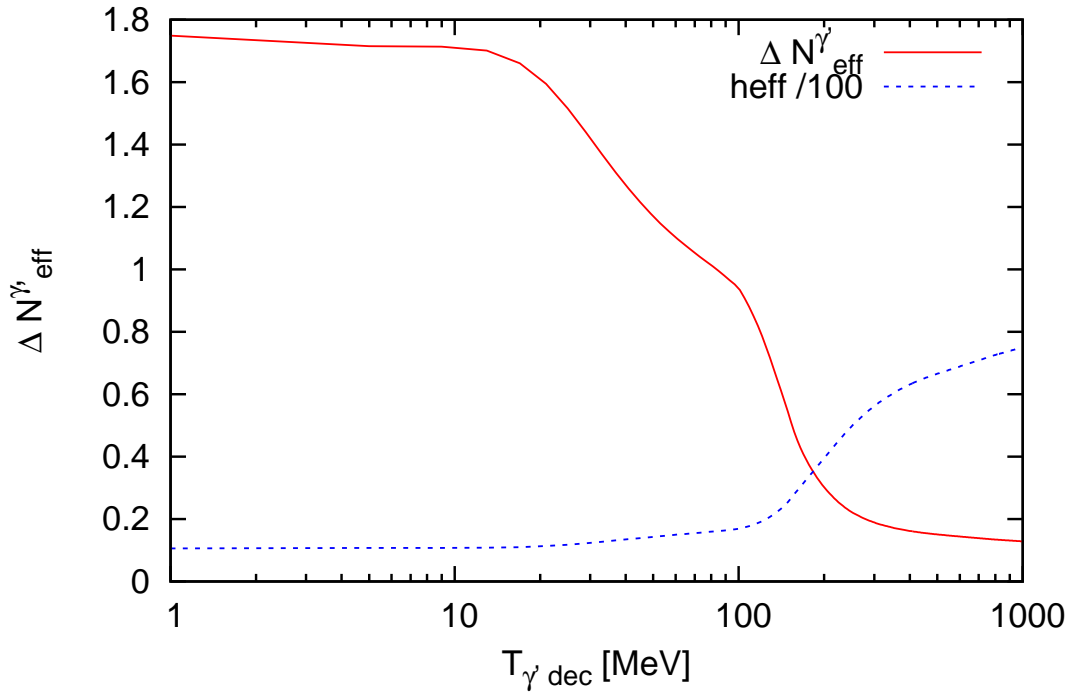}
  \caption{Dark photon contribution to the effective number of light neutrino species, 
$\Delta N^{\gapr}_{\rm eff}$,
in dependence of its decoupling temperature $T_{\gapr {\rm dec}}$ (solid line).
Also plotted is the effective degrees of freedom for the entropy density, $h_{\rm eff}$,
versus temperature $T = T_{\gapr {\rm dec}}$. 
We adopt the tabulated values assuming a QCD transition scale $T_c = 150~{\rm MeV}$ from the DarkSUSY package~\cite{Gondolo:2004sc}, and scaled them by $1/100$ (dotted line).}
  \label{fig:DNdp}
\end{figure}
\end{center}

In Fig.~\ref{fig:DNdp} we plot the dark photon contribution to the effective number of 
light neutrino species, $\Delta N^{\gapr}_{\rm eff}$,
in dependence of its decoupling temperature $T_{\gapr {\rm dec}}$.
For the effective degrees of freedom for the entropy density $\heff$,
we use the tabulated results of Ref.~\cite{Srednicki:1988ce}
from the DarkSUSY package~\cite{Gondolo:2004sc}, 
where the QCD transition scale is chosen at $T_c = 150~{\rm MeV}$.
One sees that in order to be fitted to the combined Planck + BICEP2 data, the dark photon
must decouple from the thermal bath at $T \gtrsim 100~{\rm MeV}$.
If the BICEP2 data were not included, the dark photon must decouple even ealier,
at $T \gtrsim 160~{\rm MeV}$.
As a comparison, below we also make a quick estimation based on the physical picture
as follows.
The early universe went through a rapid transition from a phase dominated by colored 
degrees of freedom (quarks and gluons) to a phase with color neutral degrees of freedom
(hadrons).
Lattice QCD calculations suggest that the transition is analytic, involving only a 
change in the dominant degrees of freedom (see e.g. Ref.~\cite{Borsanyi:2013bia}.)
Before the QCD transition we count $\heff=61.75$, which drops to 
$\heff=17.25$ after the QCD transition and before the pions become non-relativistic.
In the former case the dark photon contribution to $N_{\rm eff}$ is 
$\Delta N^{\gapr}_{\rm eff}=0.167$, while in the latter case it is 
$\Delta N^{\gapr}_{\rm eff}=0.912$.
Therefore the dark photons must decouple before the QCD transition 
in order to satisfy the CMB bound imposed by the Planck and WMAP data.
On the other hand, when the recent BICEP2 data is included, $\Delta N_{\rm eff} \sim 1$ 
is favored, and the dark photon should couple to the thermal bath until the 
$\pi^+ \pi^- \rightarrow \gamma \gamma$ annihilation.
In the next section we investigate the conditions for both scenarios.

\section{Thermal Production and Annihilation of Dark Photons}

Here we consider only the simple case:
before the QCD transition, SM particles in the thermal bath are $u$, $d$, $s$ quarks, 
gluons, muons, electrons, neutrinos, and photons.
The dark photon couples to the thermal bath mainly via the 
$\gapr \gapr \leftrightarrow \bar{s} s$, $\gapr \gapr \leftrightarrow g g$, 
and $\gapr \gapr \leftrightarrow \mu^+ \mu^-$ scattering processes.
Below the QCD transition temperature and before pions become non-relativistic, 
the dark photon couples to the thermal bath mainly via
$\gapr \gapr \leftrightarrow \pi \pi$, and 
$\gapr \gapr \leftrightarrow \mu^+ \mu^-$ scattering processes.
When the dark photon annihilation rate becomes smaller than the Hubble expansion rate at some
temperature, it freezes out.
To estimate the dark photon freeze-out temperature, in this section we calculate the 
dark photon thermally averaged annihilation cross section times the M$\o$ller velocity
\begin{equation}
   \left<\sigma_{\gapr \gapr \rightarrow F}\, v_{\rm M} \right> = 
   \frac{1}{(n^{\rm eq}_{\gapr})^2}
   \int \sigma_{\gapr \gapr \rightarrow F}\, v_{\rm M}\, 
   d n^{\rm eq}_{\gapr, 1} d n^{\rm eq}_{\gapr, 2}\, ,
\end{equation}
for all annihilation final states $F$, with 
\begin{equation}
   n^{\rm eq}_{\gapr} = \int d n^{\rm eq}_{\gapr, i} = 3\, 
   \int f (\vec{q}_i)\, \frac{d^3 \vec{q}_i}{(2 \pi)^3}\, ,
\end{equation}
the equilibrium number density of the dark photon.
Here the densities and the M$\o$ller velocity refer to the cosmic comoving frame.

{\it i}) Annihilation into quarks and muons:
the amplitude squared for $\gapr (q_1) \gapr (q_2) \rightarrow \bar{f} (p_1) f (p_2) $ is 
\begin{equation}
   \hspace{-2cm} \sum |\mathcal{M}_{\gapr \gapr \rightarrow \bar{f} f}|^2 = N_c\,
   (2 \lambda_{\Phi \chi}\, m_f)^2\, 
   \frac{4 \left[(p_1 \cdot p_2) - m^2_f \right]}{(s-m^2_1)^2 (s-m^2_2)^2}\,   
   \left[(q_1 \cdot q_2)^2 + 2 m^4_{\gapr} \right]\, ,
\end{equation}
where the center-of-mass energy is $s = (q_1 + q_2)^2 = (p_1 + p_2)^2$.
The amplitude is summed over the polarization states of the dark photons in the 
initial state and the spins of the final state fermions.
The factor $N_c$ comes from summing over final quark colors, with the color factor
$N_c = 3$ for quarks and 1 for leptons.
One sees that at energies $\sqrt{s} \gg m_{\gapr}$, the Goldstone boson contribution
dominates over that from the transverse polarization state of the dark photon.
In the large $m_2$ limit, the propagator of the dark Higgs can be expanded in powers of 
$s/m^2_2$. 
In this work we used only the leading term in the expansion, $1/m^4_2$, as in 
Ref.~\cite{Weinberg:2013kea}.
The results we will present should thus be regarded as conservative estimates, 
since all higher terms contribute positively to the dark photon collision 
rate~\cite{Keung:2013mfa}.

The annihilation cross section is 
\begin{equation}
   \hspace{-2cm} \sigma_{\gapr \gapr \rightarrow \bar{f} f}  (s) = 
   \frac{1}{2 \omega_1 2 \omega_2\, v_{\rm M}}\, \frac{1}{8 \pi}\, 
   \left( \frac{1}{3} \right)^2\, \sqrt{1 - \frac{4 m^2_f}{s}}\, 
   \sum |\mathcal{M}_{\gapr \gapr \rightarrow \bar{f} f}|^2\, ,
\end{equation}
where $\omega_1, \omega_2$ denote the dark photon energies.
The M$\o$ller velocity is defined by
\begin{equation}
   v_{\rm M}\, \omega_1 \omega_2 = \sqrt{(q_1 \cdot q_2)^2 - m^4_{\gapr}}\, .
\end{equation}

\begin{center}
\begin{figure}[h!]
  \includegraphics[width=0.6\textwidth,angle=-90]{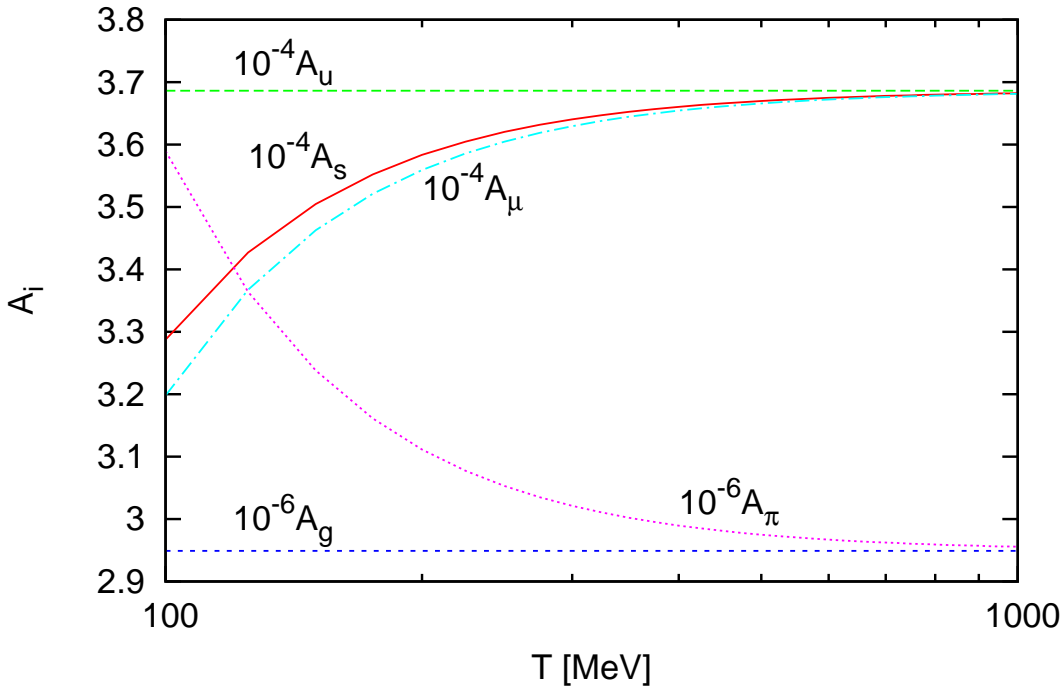}
  \caption{Temperature dependence of the $A_f$, $A_g$, and $A_\pi$ integrals, which are 
scaled by $10^{-4}$, $10^{-6}$ and $10^{-6}$ in this figure, respectively.
Among the $A_f$ integrals, we plot that for the up quark $(A_u)$, and the strange 
quark $(A_s)$, as well as that for the muon $(A_\mu)$.}
  \label{fig:Ai}
\end{figure}
\end{center}

Using Maxwell-Boltzmann statistics for the dark photons,
the thermally averaged annihilation cross section times the M$\o$ller velocity 
can be reduced to the simple one-dimensional integral~\cite{Gondolo:1990dk,Edsjo:1997bg} 
\begin{equation}
   \hspace{-2cm} \left<\sigma_{\gapr \gapr \rightarrow f \bar{f}}\, v_{\rm M} \right> = 
   \frac{3^2}{(n^{\rm eq}_{\gapr})^2}\, \frac{2 \pi^2}{(2 \pi)^6}\, T\, 
   \int^\infty_{4 m^2_{\gapr}} d s\, \sigma_{\gapr \gapr \rightarrow f \bar{f}}\, 
   (s - 4 m^2_{\gapr})\, \sqrt{s}\, K_1 \left( \frac{\sqrt{s}}{T} \right)\, ,
   \label{thermalsigmav}
\end{equation}
with $K_1 (z)$ the modified Bessel function of the second kind of order 1.
Changing to the dimensionless variables 
$u \equiv s /T^2$, $v \equiv 4 m^2_{\gapr} / T^2$, and $w \equiv 4 m^2_f / T^2$,
one can rewrite the above expression as  
\begin{equation}
   \hspace{-2cm} \left<\sigma_{\gapr \gapr \rightarrow f \bar{f}}\, v_{\rm M} \right> = 
   \frac{T^{10}}{(n^{\rm eq}_{\gapr})^2}\, \frac{\pi}{(2 \pi)^6}\,  
   \frac{N_c}{4}\, \left(\frac{\lambda_{\Phi \chi}\, m_f}{m^2_1 m^2_2} \right)^2
   \cdot A_f\, ,
\end{equation}
with 
\begin{equation}
   A_f = \int^\infty_v d u\, \sqrt{1 - \frac{w}{u}}\, (u-w)\, (u^2 - u v + \frac{3}{4} v^2)\,
   \sqrt{u-v}\, K_1 (\sqrt{u})\, .
\end{equation}
In Fig.~\ref{fig:Ai} we plot $A_f$ for $f=u$ and $s$ quarks as well as for $\mu^\pm$ in
the temperature range $T = 100$-$1000~{\rm MeV}$,
where we have set $m_{\gapr}=1~{\rm eV}$.
For the light quarks and leptons, $A_f$ is temperature independent, while 
for the heavy ones there is the mass threshold effect.

{\it ii}) Annihilation to gluons and photons:
the amplitude for $\gapr (q_1) \gapr (q_2) \rightarrow g (p_1) g (p_2)$ is
\begin{eqnarray}
   \sum |\mathcal{M}_{\gapr \gapr \rightarrow g g}|^2 &=& 8
   (2 \lambda_{\Phi \chi}\, v)^2\, 
   \frac{\left[(q_1 \cdot q_2)^2 + 2 m^4_{\gapr} \right]}{(s - m^2_1)^2 (s - m^2_2)^2}\,   
   \left(\frac{\alpha_s}{4 \pi} \right)^2\, \frac{8\, G_F}{\sqrt{2}}\,
   |F_g|^2\, \cdot 2\, (p_1 \cdot p_2)^2\, , 
\end{eqnarray}
where $G_F / \sqrt{2} = g^2_2 / (8\, m^2_W)$ is the Fermi constant.
The strong coupling constant $\alpha_s (Q)$ runs from 0.35 at $Q=2~{\rm GeV}$ down smoothly
to 0.118 at $Q= m_Z$~\cite{Beringer:1900zz}, where $Q=\sqrt{s}$ is the momentum transfer in the virtual Higgs decay process.
The form factor $F_g (Q^2 = s)$ receives contributions from all quarks.
One can approximate it with $|F_g| \rightarrow (2/3)\, n_H$, with $n_H$ the number 
of heavy quark flavors with masses $\gg \sqrt{s}$~\cite{Leutwyler:1989tn,Gunion:1989we}.
The amplitude squared is summed over the initial and the final polarization states.
Note that at temperature $T$, to do the thermal averaging one needs to integrate over the energy range $1 \lesssim \sqrt{s} / T \lesssim 20$.
However, for the sake of simplicity here we do not integrate the $F_g$ and $\alpha_s$ 
over $s$, but take their average value in the integration range.
Therefore we have
\begin{equation}
   \hspace{-2cm} \left<\sigma_{\gapr \gapr \rightarrow g g}\, v_{\rm M} \right> = 
   \frac{T^{12}}{(n^{\rm eq}_{\gapr})^2}\, \frac{\pi}{(2 \pi)^6}\,  
   \left(\frac{1}{2} \right)\, 
   \left(\frac{\lambda_{\Phi \chi}}{m^2_1 m^2_2} \right)^2\, 
   \left(\frac{\alpha_s}{4 \pi} \right)^2\, |F_g|^2
   \cdot A_g\, ,
\end{equation}
where  
\begin{equation}
   A_g = \int^\infty_v d u\, (u^2 - u v + \frac{3}{4} v^2)\, u^2 \sqrt{u - v}\, 
   K_1 (\sqrt{u})\, .
\end{equation}
Numerically, $A_g$ is nearly constant in the whole temperature range 
$T = 100$-$1000~{\rm MeV}$, 
as shown by the blue-dashed line in Fig.~\ref{fig:Ai}.

For the $\gapr (q_1) \gapr (q_2) \rightarrow \gamma (p_1) \gamma (p_2)$ annihilation 
process, the result can be obtained by multiplying the above result by a factor of
$\frac{1}{8}\, (\alpha/ \alpha_s)^2\, |F_\gamma|^2 / |F_g|^2$.
The photon form factor at low energies is 
$|F_\gamma|^2 \simeq \mathcal{O} (1)$, with a peak and a dip stemming from
the threshold singularities generated by the $\pi^+ \pi^-$ and the $K^+ K^-$ cuts,
respectively~\cite{Leutwyler:1989tn}.

{\it iii}) Annihilation into pions:
the coupling of the SM Higgs to pions 
is~\cite{Vainshtein:1980ea,Voloshin:1985tc,Gunion:1989we} 
\begin{equation}
   \left< \pi^+ \pi^- \Big\arrowvert \mathcal{L}_{\rm int} \Big\arrowvert h \right> 
   \simeq - \frac{2}{9 v}\, \left( Q^2 + \frac{11}{2}\, m^2_\pi \right)\, ,
\end{equation}
for 3 heavy quark flavors, where $Q^2$ is the momentum transfer.
The amplitude squared for $\gapr (q_1) \gapr (q_2) \rightarrow \pi^+ (p_1) \pi^- (p_2)$ 
is then
\begin{equation}
   \hspace{-2.2cm} \sum |\mathcal{M}_{\gapr \gapr \rightarrow \pi^+ \pi^-}|^2 = 
   \left(\frac{2}{9} \right)^2 (2 \lambda_{\Phi \chi})^2\, 
   \frac{\left(s + \frac{11}{2}\, m^2_\pi\right)^2}{\left(s-m^2_1\right)^2 \left(s-m^2_2\right)^2}\, 
   \left[(q_1 \cdot q_2)^2 + 2 m^4_{\gapr} \right]\, ,
\end{equation}
which is summed over the polarization states of the initial dark photons.
We calculate the thermally averaged annihilation cross section times the M$\o$ller velocity 
\begin{equation}
   \hspace{-2cm} \left<\sigma_{\gapr \gapr \rightarrow \pi^+ \pi^-}\, v_{\rm M} \right> = 
   \frac{T^{12}}{( n^{\rm eq}_{\gapr})^2}\, \frac{\pi}{(2 \pi)^6}\,  
   \left(\frac{1}{8} \right) \left(\frac{2}{9} \right)^2\,
   \left(\frac{\lambda_{\Phi \chi}}{m^2_1 m^2_2} \right)^2 \cdot A_\pi\, ,
\end{equation}
where 
\begin{equation} 
   A_\pi = \int^\infty_v d u\, \sqrt{1 - \frac{w}{u}}\, (u + \frac{11}{8} w)^2\, 
   (u^2 - u v + \frac{3}{4} v^2)\, \sqrt{u-v}\, K_1 (\sqrt{u})\, ,
\end{equation}
and here $w \equiv 4 m^2_\pi / T^2$.
The result for $A_\pi$ is shown by the red-dotted curve in Fig.~\ref{fig:Ai}. 
Its temperature dependence is due to the decreasing contribution of $w$ in the term $(u +11w/8)^2$ with temperature.

It has been pointed out that Higgs decay to pions may be enhanced relative to the decay
to muons due to final state interactions. 
In Ref.~\cite{Raby:1988qf} the ratio
\begin{equation}
   R_{\pi \mu} \equiv \frac{\Gamma_{h \rightarrow \pi^+ \pi^- + \pi^0 \pi^0}}
   {\Gamma_{h \rightarrow \mu^+ \mu^-}}
\end{equation}
has been calculated in the Higgs mass range between $2 m_\pi$ and $2 m_K$, with $m_K$
the kaon mass.

\section{Freeze Out of Dark Photons}

Using Maxwell-Boltzmann statistics for all species, the Boltzmann equation for the 
dark photon is 
\begin{equation}
  \frac{d n_{\gapr}}{d t}  + 3 H\, n_{\gapr} = 
   - \sum_{F} \left<\sigma_{\gapr \gapr \rightarrow F}\, v_{\rm M} \right>\, 
   \left(n^2_{\gapr} - (n^{\rm eq}_{\gapr})^2 \right) 
   \equiv - \left<\sigma_{\rm ann} v_{\rm M} \right>\, 
   \left(n^2_{\gapr} - (n^{\rm eq}_{\gapr})^2 \right)\, ,
\end{equation}
where $F$ denotes all dark photon annihilation final states.
Note that the cross section appearing here is the usual one: summed over initial and final
spins, averaged over initial spins, with no factor of $1/2!$ for identical initial particles.
In the radiation-dominating epoch, the Hubble expansion rate is
\begin{equation}
   H (T) = \left(\frac{8}{3} \pi\, G_N\, \rho_R \right)^{1/2}
   \simeq 1.66\, \geffsr (T)\, \frac{T^2}{m_{\rm Pl}}\, ,
\end{equation}
where $G_N$ is Newton's constant, and $m_{\rm Pl} = G^{-1/2}_N$ the Planck mass.
It is convenient to write the Boltzmann equation in terms of the variables
$Y_{\gapr} \equiv n_{\gapr}/s$ and $x=m_{\gapr} / T$ 
as (see e.g. Ref.~\cite{Gondolo:1990dk})
\begin{equation}
   \frac{x}{Y^{\rm eq}_{\gapr}} \frac{d Y_{\gapr}}{d x} = \frac{1}{3 H} \frac{d s}{d x}\, 
   Y^{\rm eq}_{\gapr} \left<\sigma_{\rm ann} v_{\rm M} \right>\, 
   \left[ \left( \frac{Y_{\gapr}}{Y^{\rm eq}_{\gapr}} \right)^2 - 1 \right]
   = - \left( \frac{\gastsr\, \geffsr}{\heff} \right)\,
   \frac{\Gamma_{\rm ann}}{H} 
   \left[ \left( \frac{Y_{\gapr}}{Y^{\rm eq}_{\gapr}} \right)^2 - 1 \right]\, .
\end{equation}
Here $Y_{\gapr}$ and $Y^{\rm eq}_{\gapr} = n^{\rm eq}_{\gapr} / s$ are the actual and the equilibrium number of dark photon per comoving volume, respectively, and 
$\Gamma_{\rm ann} = n^{\rm eq}_{\gapr} \left<\sigma_{\rm ann} v_{\rm M} \right>$ is the 
dark photon annihilation rate.
The new degrees of freedom parameter introduced here is defined as~\cite{Gondolo:1990dk}
\begin{equation}
   \gastsr \equiv \frac{\heff}{\geffsr}\, 
   \left(1 + \frac{1}{3}\, \frac{T}{\heff} \frac{d \heff}{d T} \right)\, .
\end{equation}
One sees that when the ratio $\Gamma_{\rm ann} / H$ becomes less
than order unity, the relative change in the dark photon number 
$\Delta Y_{\gapr} / Y_{\gapr} \sim (x d Y_{\gapr} /d x)/Y^{\rm eq}_{\gapr}$ becomes small.
The number of dark photons in a comoving volume, 
$Y_{\gapr}$, freezes in and starts to deviate from its equilibrium value 
$Y^{\rm eq}_{\gapr}$.
Without solving the Boltzmann equation numerically, in this work
we determine the freeze-out temperature of the dark photons by requiring
\begin{equation}
\label{eq:frout_criterion}
   \left( \frac{\gastsr\, \geffsr}{\heff} \right)\, \frac{\Gamma_{\rm ann}}{H}
   \simeq 1\, ,
\end{equation}
at $T = \tdpdec$.

Since the QCD transition is not a real phase transition, but only involves a change in
the dominant degrees of freedom, there is no uniquely defined transition temperature
$T_c$~\cite{Borsanyi:2013bia}.
Generally $T_c$  lies in the range 
$150 - 170~{\rm MeV}$~\cite{Borsanyi:2010bp,Tawfik:2011sh,Bazavov:2011nk}.
Furthermore, lattice calculations indicate that the quark-gluon plasma can be described by 
free quarks and gluons only for $T \gtrsim 4\, T_c$~\cite{Boyanovsky:2006bf}.
However, in this work we derive bounds on the dark sector parameters assuming
the validity of the free quark and gluon picture for all temperatures above $T_c$.
For the $\gastsr (T)$ and $\heff (T)$ functions, we use the results obtained 
in Ref.~\cite{Srednicki:1988ce} which were tabulated in the DarkSUSY 
package~\cite{Gondolo:2004sc}, where $T_c = 150~{\rm MeV}$ was chosen.

\subsection{Above the QCD Transition Temperature}

The total thermally averaged annihilation cross section times the M$\o$ller velocity is
\begin{equation}
   \left<\sigma_{\rm ann}\, v_{\rm M} \right> = 
   \sum_V \left<\sigma_{\gapr \gapr \rightarrow V V}\, v_{\rm M} \right> + 
   \sum_f \left<\sigma_{\gapr \gapr \rightarrow f \bar{f}}\, v_{\rm M} \right>\, ,
\end{equation}
with $V = g, \gamma$, and $f = u, d, s, \mu^{\pm}, e^{\pm}$.
From Eq.~(\ref{eq:frout_criterion}),
the dark photon freezes out at temperature $\tdpdec$ such that
\begin{equation} 
\label{eq:aboveQCD}
   \frac{1}{192\, \zeta (3) \pi^3}
   \left( \frac{\lambda^2_{\Phi \chi}\, T^7_{\dpdec}}{m^4_1 m^4_2} \right)
   \left[\left(\frac{\alpha_s}{4 \pi} \right)^2 |F_g|^2\, \frac{A_g}{2}\,
   T^2_{\dpdec} + \left( \sum_f N_c\, m^2_f\, \frac{A_f}{4} \right) \right] 
   \simeq 1.66\, \frac{\heff}{\gastsr}\, 
   \frac{T^2_{\dpdec}}{m_{\rm Pl}}\, ,
\end{equation}
where the first term is the contribution from the gluon, and the second term that from 
the quark ($N_c = 3$) and lepton ($N_c = 1$) channels.
Here $\zeta (3) \approx 1.202$ is the Riemann zeta function $\zeta(z)$ evaluated at $z=3$.
The photon channel contribution is only $\sim 10^{-5}$ times that from gluon channel, 
totally negligible.
We approximate the gluon form factor with $|F_g| \sim 2$., and 
the strong coupling constant $\alpha_s (Q) \sim 0.2$. 
Our results for the dark photon freeze-out conditions are displayed in 
Fig.~\ref{fig:lambda_bound}, where we have assumed the validity of 
Eq.~(\ref{eq:aboveQCD}) down to $\tdpdec=150~{\rm MeV}$.
As an example, we find
\begin{equation}
   \lambda_{\Phi \chi} \simeq
   2.85 \times 10^{-3}\, \left(\frac{m_2}{1~{\rm GeV}} \right)^2\, ,
\end{equation}
for $\tdpdec = 200~{\rm MeV}$, and
\begin{equation}
   \lambda_{\Phi \chi} \simeq
   1.02 \times 10^{-4}\, \left(\frac{m_2}{1~{\rm GeV}} \right)^2\, ,
\end{equation}
for $\tdpdec = 700~{\rm MeV}$.
The latter scenario is not constrained by current collider sensitivites to the invisible 
decay width of the SM Higgs, unless the dark Higgs is as heavy as $\sim 10~{\rm GeV}$.
The ILC with the projected sensitivity as mentioned in Sec.~III would have a chance to 
probe this scenario if $m_2$ is larger than $4~{\rm GeV}$.

\subsection{Below the QCD Transition Temperature}

\begin{center}
\begin{figure}[t!]
  \includegraphics[width=0.6\textwidth,angle=-90]{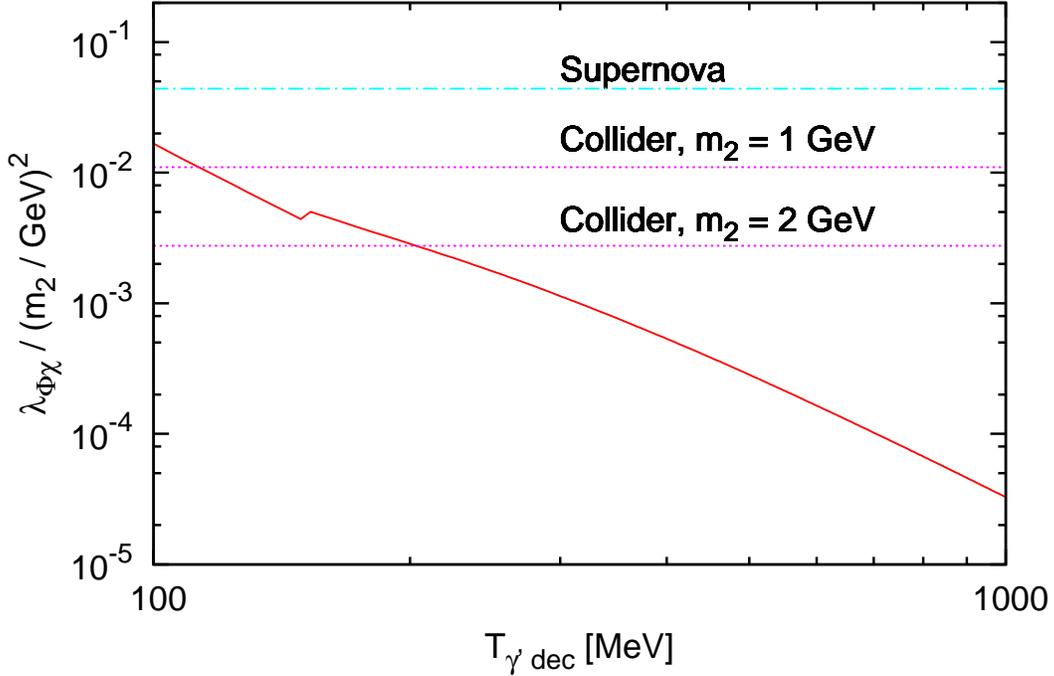}
  \caption{Required value for the Higgs portal coupling $\lambda_{\Phi \chi}$ scaled with the mass squared of the dark Higgs $m_2$ in units of GeV, in order that the 
dark photon freezes out at temperature $\tdpdec$ (solid line). 
Also shown are the supernova bound (dash-dotted), as well
as the collider bounds for $m_2 = 1~{\rm GeV}$ and $2~{\rm GeV}$ (dotted). 
Each horizontal line excludes the region above it.}
  \label{fig:lambda_bound}
\end{figure}
\end{center}

The total thermally averaged annihilation cross section times the M$\o$ller velocity is
\begin{equation}
   \left<\sigma_{\rm ann}\, v_{\rm M} \right> = 
   \left<\sigma_{\gapr \gapr \rightarrow \pi \pi}\, v_{\rm M} \right> + 
   \sum_f \left<\sigma_{\gapr \gapr \rightarrow f \bar{f}}\, v_{\rm M} \right> + 
   \left<\sigma_{\gapr \gapr \rightarrow \gamma \gamma}\, v_{\rm M} \right>\, ,
\end{equation}
with $f = \mu^{\pm}, e^{\pm}$.
Requiring $T_c > \tdpdec$,
dark photon freezes out at temperature $\tdpdec$ such that
\begin{equation}
\label{eq:belowQCD}
   \frac{1}{192\, \zeta (3) \pi^3}
   \left( \frac{\lambda^2_{\Phi \chi}\, T^7_{\dpdec}}{m^4_1 m^4_2} \right)
   \left[\frac{A_\pi}{162}\, T^2_{\dpdec} + \sum_f m^2_f \frac{A_f}{4} \right] \simeq 
   1.66\, \frac{\heff}{\gastsr}\, \frac{T^2_{\dpdec}}{m_{\rm Pl}}\, ,
\end{equation}
where the first and the second term is the pion and the muon contribution, respectively.
The results are displayed in Fig.~\ref{fig:lambda_bound}. 
The kink at $\tdpdec=150~{\rm MeV}$ arises from the mismatching of the two 
freeze-out criteria in Eqs.~(\ref{eq:aboveQCD}) and (\ref{eq:belowQCD}) at this point.
The above condition is translated to 
\begin{equation}
   \lambda_{\Phi \chi} \simeq 0.0054\, \left(\frac{m_2}{1~{\rm GeV}} \right)^2\, ,
\end{equation}
if $\tdpdec = 140~{\rm MeV}$, and
\begin{equation}
   \lambda_{\Phi \chi} \simeq 0.0167\, \left(\frac{m_2}{1~{\rm GeV}} \right)^2\, ,
\end{equation}
for $\tdpdec = 100~{\rm MeV}$.
One sees that the current collider bound (Eq.~(\ref{colliderbound})) requires that 
$m_2 \lesssim 1~{\rm GeV}$ in order that the dark photon plays the role of fractional neutrinos and contributes roughly $0.9$ to $N_{\rm eff}$.

\section{Supernova Bound}

The observed duration of neutrino burst events from Supernova 1987A in several detectors 
confirmed the standard picture of neutrino cooling of post-collapse supernova.
In the second phase of neutrino emission, a light particle which interacts even more weakly than neutrinos could lead to more efficient energy loss and shorten the neutrino burst duration.
Demanding that the novel cooling agent $X$ should not have affected the  
total cooling time significantly,
an upper bound on their emissivity can be derived~\cite{Raffelt:1990yz,Raffelt:2006cw}
\begin{equation}
\label{emissivity_bound}
   \epsilon_X \equiv \frac{Q_X}{\rho} \lesssim 10^{19}\, {\rm erg} \cdot {\rm g}^{-1} \cdot
  {\rm s}^{-1} = 7.324 \cdot 10^{-27}\, {\rm GeV}\, ,
\end{equation}
where $Q_X$ is the energy loss rate and $\rho$ is the core density.
This bound, dubbed  ``Raffelt criterion", is to be applied at typical core conditions, i.e. a mass 
density $\rho = 3 \cdot 10^{14}$ g/cm$^3$ and a temperature $T = 30$ MeV.
The self-consistent cooling calculations and statistical analysis
performed for the Kaluza-Klein gravitons in Ref.~\cite{Hanhart:2001fx} 
demonstrated the reliability of this simple criterion.

In the gauged Weinberg's Higgs portal model, dark photon pairs can be produced in
the $e^+ e^- \rightarrow \gapr \gapr$, $\gamma \gamma \rightarrow \gapr \gapr$ annihilation
processes, and most efficiently in the nuclear bremsstrahlung processes 
$N N \rightarrow N N \gapr \gapr$ in a post-collapse supernova core.
After the production, free-streaming of individual dark photon out of the core may lead 
to significant energy loss rate, depending on the portal coupling, the dark Higgs mass, as
well as the coupling of the SM Higgs to the nucleons.
Since the energy of the emitted dark photons is 
of the order of the core temperature
which is considerably larger than its mass in the present consideration, 
one can appeal to the equivalence theorem to deduce a bound from the Goldstone boson calculation~\cite{Keung:2013mfa},
\begin{equation}
   \lambda_{\Phi \chi} \lesssim 0.044 \left(\frac{m_2}{1~{\rm GeV}} \right)^2\, ,
\end{equation}
in the large $m_2$ limit.
The supernova bound is comparable to collider bounds in the case of 
$m_2 \lesssim 500~{\rm MeV}$, and is generally weaker than those derived from the 
freeze-out criterion (cf. Fig.~\ref{fig:lambda_bound}).

\section{Summary and Outlook}

In this work we have investigated the viability of the dark photon arising from gauged Weinberg's Higgs portal model of playing the role of fractional cosmic neutrinos.
If the dark photon decouples from the thermal bath after the QCD transition 
and before pions become non-relativistic, it contributes to the effective number of neutrino species $N_{\rm eff}$ with $\sim 0.9$, compatible with what inferred by the recent BICEP2
B-mode polarization data.
We estimated the dark photon freeze-out temperature in dependence of 
the dark Higgs mass $m_2$ and the portal coupling $\lambda_{\Phi \chi}$ between the dark and SM Higgs fields
in the large {\color{red}$m_2$} approximation.
The supernova bound on the portal coupling is the same as in the 
Goldstone boson case, thus being an order of magnitude weaker than those derived from the 
freeze-out criteria.
Combining with the Higgs invisible width constraint obtained from the global fits from the latest LHC data, we find that the dark Higgs boson mass is required to be lighter than about 
$1.5~{\rm GeV}$.
In the future, a projected sensitivity of the ILC to Higgs invisible decay may strengthen
this bound by a factor of 2.6.
On the other hand, if future CMB observations are in favor of a smaller $N_{\rm eff}$,
the dark photon has to decouple before the QCD transition. 
In this case the portal coupling $\lambda_{\Phi \chi}$ is getting smaller and the dark Higgs mass $m_2$ is less constrained by the colliders. 

In summary, the original abelian $U(1)$ Higgs model is used as a dark sector and we have studied some interesting implications in both the early universe as well as TeV collider physics. 
Perhaps this original toy model for spontaneous symmetry breaking in relativistic quantum field theories can be realized in the invisible world.


\begin{acknowledgments}
This work was supported in part by the Ministry of Science and Technology, Taiwan,
ROC under the Grant Nos. 101-2112-M-001-010-MY3  (KWN, HT) and
101-2112-M-001-005-MY3 (TCY).
\end{acknowledgments}

\end{document}